\documentclass{aa}

\usepackage{graphicx}
\usepackage{xcolor}
\usepackage{txfonts}
\usepackage[draft]{hyperref}
\usepackage{placeins}

\newcommand{\ngdor}{28}

\newcommand{\nhybrid}{19}
\newcommand{\gdor}{$\gamma$\,Dor}
\newcommand{\dscu}{$\delta$\,Sct}

\begin{document}
    \title{Probing stellar rotation in the Pleiades with gravity-mode pulsators
    }


    \author{D. J. Fritzewski\inst{1}
        \and
        A. Kemp\inst{1}
        \and
        G. Li\inst{1,2}
        \and
        C. Aerts\inst{1,3,4}
    }

    \institute{Institute of Astronomy, KU Leuven, Celestijnenlaan 200D, 3001, Leuven, Belgium\\
        \email{dario.fritzewski@kuleuven.be}
        \and
        Centre for Astrophysics, University of Southern Queensland, Toowoomba, QLD 4350, Australia
     \and Department of Astrophysics, IMAPP, Radboud University Nijmegen, PO Box 9010, 6500 GL Nijmegen, The Netherlands
      \and Max Planck Institut für Astronomie, Königstuhl 17, 69117 Heidelberg, Germany
    }

    \date{}



    \abstract
    {Due to their proximity, the Pleiades are an important benchmark open cluster. Despite its status, asteroseismic analyses of its members are rare. In particular, the gravity-mode (g-mode) pulsators, which allow inference of stellar near-core properties have not been analysed yet.}
    {We aim to identify and analyse the population of g-mode pulsators in the Pleiades. Our focus lies on the internal rotation as measured from asteroseismology to obtain a well defined sample of stellar rotation on the early main sequence.}
    {Based on full-frame images from the Transiting Exoplanet Survey Satellite (TESS), we constructed light curves for intermediate-mass Pleiades members and searched for g-mode pulsators among them. For pulsators exhibiting period spacing patterns, we determined their near-core rotation rate and buoyancy periods. For all other g-mode pulsators, we estimated the near-core rotation rate based on the dominant mode frequency to obtain a comprehensive rotation rate distribution.}
    {Among our 105 target stars, we find \ngdor{} g-mode pulsators distributed across the entire upper main sequence, \nhybrid{} of which are hybrid pulsators, but only three stars exhibit period spacing patterns in the current TESS data. The near-core rotation rates in A- and early F-type members are distributed between 1 and 3\,d$^{-1}$ without any clear mass-dependence. This distribution is much broader than the one in  the similar open cluster NGC\,2516. A comparison of the buoyancy periods shows that the Pleiades and NGC\,2516 are of similar asteroseismic age.}
    {With the large population of g-mode and hybrid pulsators, the Pleiades constitute a valuable asteroseismic benchmark cluster, reaffirming its important role in stellar astrophysics.}


    \titlerunning{Rotation of gravity-mode pulsators in the Pleiades}
    \authorrunning{Fritzewski et al.}

    \maketitle
    %
\nolinenumbers
    \section{Introduction}
    From pre-historic times to the present day, the \object{Pleiades} open cluster has sparked human imagination. Despite being one of the most extensively studied open clusters, its stars continue to offer new insights \citep[e.g.][]{Rebull2016, Bedding2023} while simultaneously serving as calibrators in stellar evolution \citep[e.g.][]{2012MNRAS.424.3178B,2019AJ....158...77C, 2020ApJ...891...29S, Fritzewski2020}. One observational window that opened widely in the past decade is asteroseismology \citep[see reviews by][]{2020FrASS...7...70B, 2021RvMP...93a5001A, 2022ARA&A..60...31K}, leading to deeper understanding of stellar interior physics such as  stellar structure, internal rotation, and mixing profiles \citep[e.g.][]{2016A&A...593A.120V, 2018A&A...618A..24V, 2021NatAs...5..715P, 2022ApJ...940...49P,2023A&A...677A..63M}. The Pleiades have long been a crucial anchor point for stellar evolution models \citep[e.g.][]{1956ApJ...123..267J, 1957ApJ...125..435S} and are thus of particular importance for asteroseismic exploration.

    Gravity-mode (g-mode) pulsations observed in early-type stars provide information about the stellar region in the transition zone between the convective core and the radiative envelope, including its rotation rate. Such modes are therefore of great interest to efforts seeking to understand angular momentum transport in these stars \citep{2024A&A...692R...1A}. The high-order gravity modes in the lowest-mass g-mode pulsators, the $\gamma$\,Doradus (\gdor{}) stars, but also in the higher-mass slowly pulsating B (SPB) stars, are typically excited in consecutive radial orders, forming a period spacing pattern that can be exploited asteroseismically \citep{2008MNRAS.386.1487M,2013MNRAS.429.2500B,2015ApJS..218...27V}.

    The impressive potential of g-mode asteroseismology was highlighted by the \emph{Kepler} mission \citep{KeplerMission} which enabled the detailed pulsation analyses of individual stars \citep[e.g.][]{2016A&A...593A.120V}. Further, it enabled the discovery and measurement of asteroseismic parameters for hundreds of g-mode pulsators \citep{Li2020, 2020MNRAS.497.4363L}. However, most of the \emph{Kepler} stars are field stars and its population of \gdor{} stars does not cover young main sequence stars \citep{2024A&A...684A.112F}. With the Transiting Exoplanet Survey Satellite \citep[TESS,][]{Ricker2014}, other samples of \gdor{} stars of a various ages became accessible \citep[e.g.][]{2022A&A...662A..82G, 2022A&A...668A.137G, Hey2024}. One challenge still remains: asteroseismic ages are model dependent. In order to calibrate stellar evolution models and to fully understand g-mode pulsators, a sample of stars with known ages, covering the zero-age main sequence (ZAMS) to the terminal-age main sequence (TAMS), is required. Such a sample can be provided by open clusters.

    Open clusters have been the target of asteroseismic exploration for a long time \citep{1971PASP...83...84D,1972ApJ...176..367B,1972ApJ...176..373B,1977MmRAS..84..101B}. However, only the recent, long-term, high-precision, high-cadence space based observations of the Transiting Exoplanet Survey Satellite \citep[TESS,][]{Ricker2014}, in concert with the precise astrometry from \emph{Gaia} \citep{2016A&A...595A...1G,2023A&A...674A...1G} allow to probe open clusters in detail with g-mode asteroseismology. 
    In \cite{Fritzewski2024}, we showed the potential of a single g-mode pulsator in an open cluster for age dating the whole stellar population. \cite{Li2024} established the population of pulsators in the rich southern open cluster \object{NGC\,2516}, which was found to host 24 \gdor{} pulsators including eleven with  period spacing patterns allowing the measurement of near-core properties.

    The Pleiades (also known as Messier\,45 and Melotte\,22), being of similar age as NGC\,2516 \citep{Fritzewski2020, Bouma2021}, are an ideal target for ensemble asteroseismology of the cluster population. Further, the cluster allows to build a comprehensive sample of young \gdor{} pulsators and to compare it with NGC\,2516 to uncover any possible cluster-to-cluster differences. Both open clusters are among the most massive ones within 500\,pc \citep{2024A&A...686A..42H}. The Pleiades as the most prominent open cluster in the night sky is located at 135\,pc, while NGC\,2516 is less prominent due to being three times further away (408\,pc). In the context of TESS photometry this means we gain access to stars in the core of the Pleiades, while the stars in the cluster core of NGC\,2516 are blended.  For a more in depth comparison between the Pleiades, NGC\,2516, and other open clusters of the same age group, we refer the reader to \cite{Fritzewski2020}.

    Previous work on pulsating stars in the Pleiades includes \cite{1972ApJ...176..367B}, who found five $\delta$ Scuti (\dscu{}) variables (intermediate-mass pressure mode (p-mode) pulsators). Ground-based time series photometry is observationally expensive, so only a few additional \dscu{} pulsators were discovered in subsequent studies \citep{1996A&A...310..831K, 1999MNRAS.309.1051K, 1999ChA&A..23..349L, 2002A&A...382..556F, 2002A&A...395..873L}. \cite{2000A&A...358..287M} discovered the first two \gdor{} pulsators in the Pleiades. Based on data from the \emph{Kepler}/K2 mission, \cite{2017MNRAS.471.2882W} studied the naked-eye Pleiads\footnote{Throughout the paper, we use `Pleiad' for individual members, while `Pleiades' describes the entire cluster.} ($V\lesssim5$\,mag) with halo photometry. From the same data set, \cite{2022MNRAS.511.5718M} provided the first space-based photometry of eight known \dscu{} pulsators in the Pleiades. Recently, \cite{Bedding2023} expanded the number of \dscu{} pulsators to 36 stars based on three consecutive TESS sectors of the ecliptic plane survey and \cite{2023MNRAS.520L..53S} studied the eclipsing binary \dscu{} star HD\,23642. Equipped with additional data in the ecliptic plane from the second extended mission of TESS, we are now in a position to unambiguously identify and analyse the closely-spaced pulsation modes of \gdor{} stars in the Pleiades open cluster.

    The paper is structured as follows. In Section~\ref{sec:targetanddata}, we characterize our target sample and present the TESS data reduction workflow. Section~\ref{sec:pulsators} gives an overview of the pulsators in the Pleiades, while we provide our asteroseismic analysis in Section~\ref{sec:astero}, before exploring the rotation of the upper main sequence Pleiads in more detail in Sect.~\ref{sec:rotation}. Finally in Section~\ref{sec:N2516}, we compare our population of g-mode pulsators to that in the very similar open cluster NGC\,2516 and conclude in Section~\ref{sec:conclusions}.

    \section{Target selection and data reduction}
    \label{sec:targetanddata}
    \subsection{Membership and target selection}
\begin{figure}
        \includegraphics[width=\columnwidth]{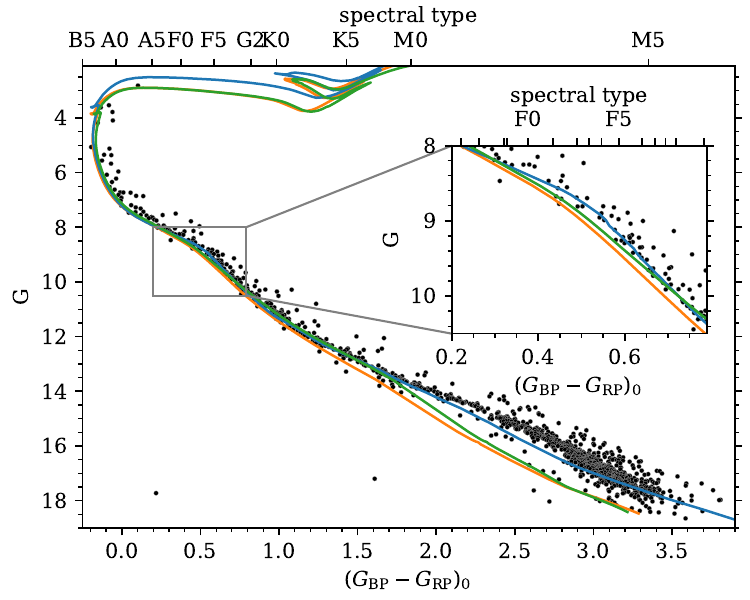}
        \caption{Colour-magnitude diagram of the Pleiades overlaid with rotating isochrones at 125\,Myr. In orange, we show \texttt{MIST} \citep{2016ApJS..222....8D, 2016ApJ...823..102C} and in blue \texttt{PARSEC} \citep{2022A&A...665A.126N}. The green isochrone is the \texttt{MIST} model with the \texttt{YBC} bolometric corrections \citep{2019A&A...632A.105C}. The inset highlights the bump around spectral type F. The top axis indicates the spectral type.
        }
        \label{fig:clusterCMD}
    \end{figure}

    Being one of the best-studied open clusters, several membership determinations of the Pleiades are available in the literature. For this work, we built on the BANYAN $\Sigma$ membership list \citep{2018ApJ...862..138G}, supplementing it with stars in the extended region around the Pleiades \citep{2021A&A...645A..84M}. The combined membership list encompasses 1213 members of the Pleiades which we show in the colour-magnitude diagram (CMD) in Fig.~\ref{fig:clusterCMD}.

    As we are interested in the variability and pulsations of stars on the upper main sequence, we selected all stars with $G<9.55$\,mag (corresponding to $M\gtrsim1.2\,M_\sun$) as our targets. This results in a sample of 105 early-type stars. The lower boundary is chosen at a relatively low mass to include all potential g-mode pulsators. Hence, we expect some of the lowest mass stars in our sample to be cool stars with surface modulations \citep{Rebull2016}, which we remove from our asteroseismic sample during the initial variability analysis.

    \subsection{Isochrones}
    The Pleiades are not only one of the best studied open clusters, but also one of the best calibration objects for stellar evolution models. Hence, we do not attempt to derive an isochronal age but instead adopt the lithium depletion age of 125\,Myr \citep{Stauffer1998}, which is in agreement with more recent age determinations \citep{2018ApJ...863...67G}. With the age fixed, we compare two different isochrones. Figure~\ref{fig:clusterCMD} shows rotating isochrones from MIST 1.2 \citep{2016ApJS..222....8D, 2016ApJ...823..102C} and \texttt{PARSEC} 2.0 \citep{2022A&A...665A.126N}. To disentangle effects stemming from the assumed colour transformations from the underlying input physics, we also include a third isochrone constructed from the \texttt{MIST} models but using the same \texttt{YBC} bolometric corrections \citep{2019A&A...632A.105C} used by the \texttt{PARSEC} isochrones.

    At first glance, the models are very similar but the accurate \emph{Gaia} photometry \citep{2021A&A...649A...3R} allows us to pin down the regions in which the isochrones stray from the observed cluster sequence.
    In this work, we focus on the intermediate-mass stars (a CMD of only this region is shown in Fig.~\ref{fig:CMD}) and we refer the reader to \cite{2025ApJ...979...92W} for a discussion of the differences between observations and models for low-mass stars.

    The \texttt{MIST} model (orange in Fig~\ref{fig:clusterCMD}, dashed in Fig.~\ref{fig:CMD}) passes close to the sub-giant Alcyone (the brightest Pleiad) and the main sequence turn-off star Electra (see Fig.~\ref{fig:CMD}). However, the majority of turn-off stars appear redder than this isochrone, suggesting different rotational properties for these stars. On the main sequence the \texttt{MIST} model mostly follows the lower envelope of the cluster sequence. However, redwards of $(G_\mathrm{BP}-G_\mathrm{RP})_0=0.35$ the isochrone deviates from the cluster sequence and predicts lower magnitudes (see inset of Fig.~\ref{fig:clusterCMD}). Using the colour transformations from \texttt{YBC}, the entire isochrone moves to slightly redder colours and closer to the majority of turn-off stars. In the deviating region of F stars the isochrone is closer to the observations and only starts to deviate from the cluster sequence for M\,dwarfs.

    The \texttt{PARSEC} model predicts a higher magnitude on the sub-giant branch than observed in Alcyone. It follows the turn-off region with similar accuracy to the \texttt{MIST} model. In the F star regime in which the \texttt{MIST} isochrone deviates from the cluster sequence, the \texttt{PARSEC} isochrone follows the data more closely but passes through the middle of the sequence rather than following its blue edge. We note that this region coincides with the mass regime in which the convective core develops.

    In conclusion, neither of the two model isochrones are able to completely describe the upper main sequence of the Pleiades at a fixed age. The small bump of the cluster sequence in the F star regime deserves further attention. In the older open cluster NGC\,6866, the roles of the two isochrones are reversed such that the observed sequence is better described by the straight \texttt{MIST} model, while the \texttt{PARSEC} isochrone is too red for those older stars (Wang et al. in prep.). This example implies that models of stellar evolution remain incomplete even at the level of CMD position. It also highlights the ground-breaking nature of \emph{Gaia} photometry. Asteroseismology is the best tool to better constrain these aspects of stellar evolution and the interior physics of the models \citep{2021RvMP...93a5001A}.

    \subsection{TESS data reduction}
    The initial 2-yr TESS mission did not observe the Pleiades cluster. Due to its position close to the ecliptic, it fell into the observing gap between the two hemispheres. However, the first extended mission included a survey of the ecliptic plane and most of our targets were observed in the sectors 42, 43, and 44. During the second extended mission, the Pleiades were revisited in sectors 70 and 71.
    This observing pattern yields five sectors with a time baseline of two years and two months, providing the frequency resolution needed to observe individual gravity modes and their period spacing patterns. Thanks to the blocks of consecutive sectors the window function is well-behaved, facilitating the identification of gravity modes.

    The data reduction follows the same workflow described in our previous works on other open clusters \citep{Li2024,Fritzewski2024}, and is based on \cite{2022A&A...662A..82G}. This analysis chain is particularly powerful to select g-mode period spacing patterns \citep{2022A&A...668A.137G}.  Here, we recall only the most important steps. We first downloaded 25\,px by 25\,px cut-outs from the TESS full frame images using the \texttt{TESScut API} \citep{2019ascl.soft05007B}. The optimal aperture, considering the background and flux contributions of neighbouring stars, was determined and we then extracted the light curve using this aperture. During the extraction, we excluded time intervals near the sector ends and mid-points that contain strong non-astrophysical signals. Subsequently, we detrended each light curve using a principal component analysis and removed residual long-term trends with a 5th-degree polynomial fit. Finally, all light curves were stitched into a single long-baseline light curve with a common cadence of 10\,min. For reasons of simplicity, we excluded the seven stars that fell into Sector 18 with a 30\,min cadence in our initial analysis. For one g-mode pulsator, we later included the Sector 18 data into our analysis.

    \section{Pulsators in the Pleiades}
    \label{sec:pulsators}
    We identified and classified all pulsating stars in the Pleiades by investigating their TESS light curves and their frequency spectra manually. During this analysis, we paid special attention to low-amplitude gravity modes in stars with strong p-mode pulsations (i.e. hybrid pulsators). In total, we found \ngdor{} stars to show potential gravity modes, including \nhybrid{} hybrid pulsators.

    \subsection{Gravity-mode pulsators}

    \begin{figure}
        \includegraphics[width=\columnwidth]{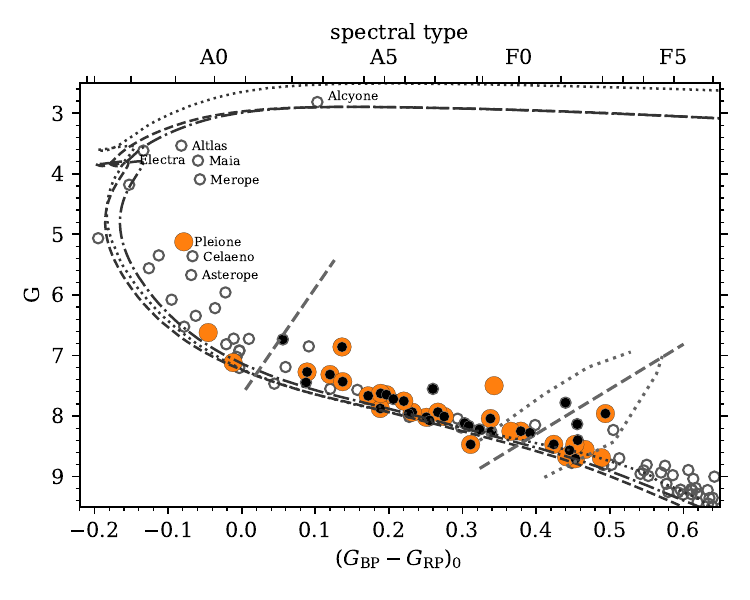}
        \caption{Colour-magnitude diagram of the upper main sequence of the Pleiades. We mark our discovered g-mode pulsators with orange and known p-mode pulsators from \cite{Bedding2023} with black dots. The dotted and dashed lines crossing the cluster sequence, indicate the \gdor{} and \dscu{} instability strips, respectively \citep{2004A&A...414L..17D, 2019MNRAS.485.2380M}. The grey lines show the isochrones from Fig.~\ref{fig:clusterCMD} with \texttt{MIST} dashed, \texttt{PARSEC} dotted, and \texttt{MIST} with the \texttt{YBC} bolometric corrections adopted in \texttt{PARSEC} dash-dotted. Stars with proper names are labelled.
        }
        \label{fig:CMD}
    \end{figure}

    Figure~\ref{fig:CMD} shows the population of g- and p-mode pulsators in a colour-magnitude diagram. As frequently observed in open clusters \citep{Li2024, Fritzewski2024} and field stars \citep{DeRidder2023, Aerts2023, Hey2024, 2024A&A...691A.131M}, the entire upper main sequence is populated with g-mode pulsators reaching from the red edge of the classical \gdor{} instability strip (spectral type  F3) to the highest mass B stars. This continuum of g-mode pulsators indicates that mode excitation occurs even outside of the well established (and theoretically described) instability region \citep{2004A&A...414L..17D,2015A&A...580L...9W}.

    Turning to the individual pulsators, we show their frequency spectra\footnote{For the accompanying light curves see Fig.~\ref{fig:lc_full} in the Appendix.} in Fig.~\ref{fig:Pgs}. Even within this homogenous population of pulsating cluster members, the signal-to-noise ratio varies strongly. Some stars show very prominent gravity modes, while the low amplitude g-mode pulsators are often hybrid pulsators with a stronger p-mode component. We investigate the population of hybrid pulsators below but first focus on individual g-mode pulsators of historical interest.

    \begin{figure}
        \includegraphics[width=\columnwidth]{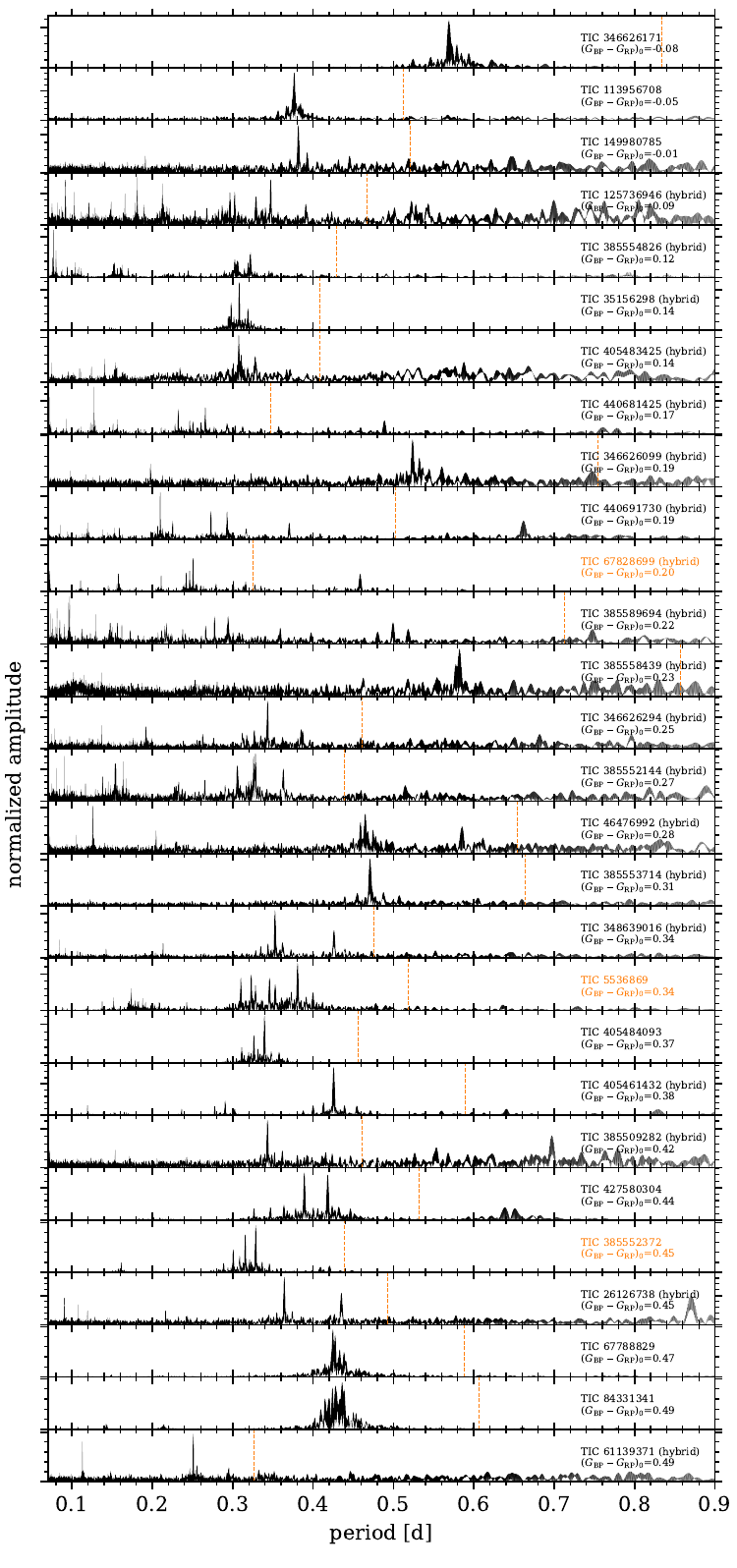}
        \caption{Periodograms of Pleiades g-mode pulsators ordered by their intrinsic colour with the hottest stars (SPB stars) at the top and the typical \gdor{} pulsators at the bottom. The orange dashed lines mark the derived near-core rotation periods. The amplitude of each periodogram is normalized by the dominant g-mode amplitude. The names of stars for which we constructed period-spacing patterns are highlighted in orange.
        }
        \label{fig:Pgs}
    \end{figure}

    \gdor{} stars were established as a class of pulsators in the 1990s \citep{1994MNRAS.270..905B, 1999PASP..111..840K} after several longer-period '\dscu{}` stars were found \citep{1986A&A...164...40A, 1993MNRAS.263..781K}. In retrospect, one can find earlier evidence of the observations of g-mode pulsators in the literature. One of these stars is HD\,23375 (TIC\,385552372). It was observed by \cite{1972ApJ...176..367B} as a variable star with a longer period than the typical \dscu{} members of the Pleiades. Despite the doubts expressed by \cite{1972ApJ...176..367B}, it entered the literature as a \dscu{} star. \cite{2000A&A...358..287M} searched the Pleiades for \gdor{} stars and observed HD\,23375 as a variable star but did not find its periodicity. The K2 observations showed longer periodic variability \citep{Rebull2016b} and \cite{Bedding2023} did not detect \dscu{} variability. Here, we confirm these findings with clear evidence of g-mode oscillations and no sign of \dscu{} variability in the TESS photometry. Hence, we can conclude that HD 23375 is a long overlooked \gdor{} star.

    HD\,22702 (TIC\,427580304) and HD\,23585 (TIC\,405484093) were the first recognized \gdor{} stars in the Pleiades \citep{2000A&A...358..287M}. Their light curves show archetypical \gdor{} variability, with clear pulsational signals in the light curve and as well as beating patterns.

    HD 23763 (TIC 35156298) is a hybrid pulsator with a dominant g-mode component. It has been described previously as a radial velocity variable star \citep{1965ApJ...142.1604A, 1975PDAO...14..319P, 1991ApJ...377..141L, 2020ApJ...901...91T} but no orbital solution for a spectroscopic binary has been found. We speculate that the observed radial velocity variability might originate from the pulsations as is commonly found in bright \gdor{} stars \citep{2004A&A...415.1079A}. This star is also included in K2 data analysis of \cite{Rebull2016b} as a multi-periodic star and classified as a \dscu{} candidate.

    With the traditional photometry as conducted above, we are not able to probe the brightest Pleiades members but only up to 5th magnitude. The brighter stars contain saturated pixels and should be treated with a halo photometry approach as in \cite{2017MNRAS.471.2882W}. Just below the magnitude cut-off, we find the brightest analysed Pleiad, Pleione (TIC\,346626171), which is a well known Be star \citep{1995PASJ...47..195H}. \cite{2017MNRAS.471.2882W} observed its variability and we confirm its pulsations. The longer time base enables the resolution of some of the closely spaced modes in its frequency spectrum. Its Be phenomenon might be caused by the combination of its fast rotation and nonlinear mode coupling among its gravity modes, as first observed in CoRoT data of Be stars \citep{Huat2009} and later found in \emph{Kepler} data of many Be \citep{2015MNRAS.450.3015K} and
    SPB \citep{VanBeeck2021} pulsators. Unfortunately, the frequency resolution of the TESS photometry is not sufficient for a detailed asteroseismic analysis.

    \subsection{Hybrid pulsators}
    Hybrid pulsators showing both g- and p-modes are of special astrophysical interest as these modes are sensitive to physics in different regions of the stellar interior. Gravity-mode pulsators are observed on the entire upper main sequence. Thus, one can expect hybrid pulsators throughout the whole \dscu{} instability strip rather than only in the small overlap between the classical \gdor{} and \dscu{} instability regions \citep{Kliapets2025}. In fact, one might even argue that all \dscu{} stars are hybrid pulsators with often undetectable (i.e. very low amplitude) gravity modes.

    Based on the \dscu{} stars in the Pleiades \citep{Bedding2023}, we identified \nhybrid{} hybrid pulsators (cf. Fig.~\ref{fig:pg_full_hybrids}) out of the 50 Pleiades members in the \dscu{} instability region. With a hybrid pulsator fraction of 38\,\%, hybrid pulsators are very abundant in the Pleiades. Moreover, 56\,\% of the \dscu{} stars and 76\,\% of the g-mode pulsators are hybrid pulsators (excluding the stars more massive than the blue-edge of the \dscu{} instability strip). These numbers are higher than corresponding counts in field star samples \citep{2016MNRAS.460.1970B, Li2020, 2022A&A...666A..76A}, making the Pleiades an ideal laboratory to study hybrid pulsators but raising questions about why so many of these pulsators are hybrid.

    One challenge in understanding hybrid pulsators comes from the difficulty in finding stars with identifiable modes for asteroseismic modelling. For gravity modes, mode identification can be achieved through modelling the behaviour of period spacing patterns, while pressure modes can be identified from their large frequency separation \citep{2020Natur.581..147B}. Within the current data set, only one hybrid pulsator (TIC\,67828699 = HD\,23388) exhibits a (short) period spacing pattern. Still, with the possibility of measuring both the period spacing and the large frequency separation or rotational multiplets in this fast rotator, it is a very promising asteroseismic target. Its detailed asteroseismic modelling will be treated in a separate, dedicated study.

    \subsection{Non-pulsators}
    \label{sec:nonpuls}
    In contrast to the hybrid pulsators, but likewise of interest to asteroseismology, are the non-pulsating stars. These cluster members might hold the key to why some stars pulsate while others do not \citep{2019MNRAS.485.2380M}. Among the analysed stars, 49 stars bluer than the red-edge of the \gdor{} instability region are not found to be pulsators. Of these stars 20 could not be analysed due to either high contamination in their light curves or being too close to the saturation limit for accurate photometry. Among the remaining 29 stars, twelve stars have sufficiently low levels of noise that pulsations with $\gtrsim 0.05$\,ppt, if present, should have been detectable. However, these stars could still pulsate but with low amplitudes which are not observable with the current data.

    Finally, 17 stars show rotational variability with periods between 0.3\,d and $\sim\!5$\,d. Two of these stars are found at the red edge of the \gdor{} instability region and could be genuine cool star rotators. Two additional stars are found within the \dscu{} instability region while the majority are higher mass stars.

    \section{Asteroseismic analysis}
    \label{sec:astero}
    Gravity modes provide a window into the stellar near-core region, with probing power in rotation, stellar structure, and chemical composition gradients \citep{Aerts2010}. Pulsation modes of the same degree (most often we observe prograde dipole modes, $l=1$, $m=1$; \citealt{Li2020}) exhibit a comb of pulsations modes that are nearly equidistant in period for consecutive radial orders. Based on this comb of frequencies, we constructed period spacing patterns, which are shown as the difference of neighbouring modes against their mode period (see Fig.~\ref{fig:PSP}). The pattern's slope is related to the near-core rotation frequency, $f_\mathrm{rot}$ \citep{2016A&A...593A.120V}, while its spacing is related to the asymptotic buoyancy period, $\Pi_0$ \citep{2022A&A...668A.137G}.
    This buoyancy period is proportional to the inverse of the integral over the Brunt-Väisälä frequency, $N$, normalized by the local radius, $r$, integrated
    over the ``g-mode cavity (gc)'',
    $\Pi_0\,\equiv\,2\pi^2\left[\int_{\rm gc} (N/r) {\rm d}r\right]^{-1}$. It encodes stellar near-core properties in one number. Structural features in the near-core region such as chemical gradients perturb the regularity of the pattern and create oscillatory patterns \citep{2018A&A...614A.128P, 2021A&A...650A.175M, 2021A&A...650A..58M}.

    Our time series are relatively short for g-mode asteroseismology and many of our g-mode pulsators do not show identifiable patterns, while those that do are very short. Moreover, the patterns are often non-consecutive due to missing modes, hindering a clear identification.

    \subsection{Period spacing pattern}
    To construct the period spacing patterns, we extracted all significant frequencies from the observations using the pre-whitening code \texttt{STAR SHADOW}\footnote{\url{https://github.com/LucIJspeert/star_shadow}} \citep{starshadow, starshadow_ascl}. We kept the default parameters and applied it to all g-mode pulsators in Fig.~\ref{fig:Pgs}.

\begin{figure*}
        \includegraphics[width=\textwidth]{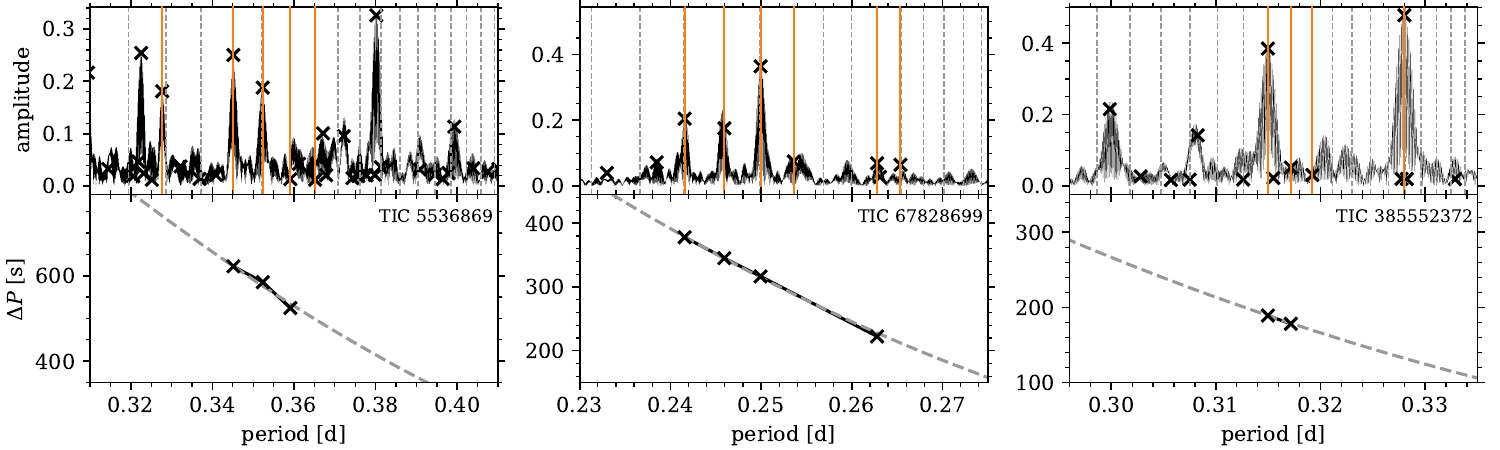}
        \caption{Period spacing patterns for the three Pleiads with regularly-spaced gravity modes. The top panels show the frequency spectra with the extracted periods marked by crosses. The selected modes of the period spacing pattern are shown by solid orange lines. The best-fit asymptotic pulsation solution is indicated by dashed lines. The lower panel shows the observed period spacing pattern (crosses and solid line) together with the best fit (dashed) for $f_\mathrm{rot}$ and $\Pi_0$.}
        \label{fig:PSP}
    \end{figure*}

    The cleanest g-mode pulsators often exhibit two combs of gravity modes in Fig.~\ref{fig:Pgs} that can be associated with $l=1$ and $l=2$ ridges (e.g. V1225 Tau = TIC 427580304). The combs of the latter modes always have a lower amplitude compared with the former. Despite these obvious g-mode combs, most of our identified g-mode pulsators do not exhibit clear patterns due to missing low amplitude modes. We find candidate period spacing patterns for three stars out of the \ngdor{} g-mode pulsators (Fig.~\ref{fig:PSP}). Interestingly, these three do not include the two previously identified \gdor{} stars \citep{2000A&A...358..287M} which host dominant gravity modes.

    We analysed each of the identified patterns with the open source package \texttt{AMiGO}\footnote{\url{https://github.com/TVanReeth/amigo}} \citep{2016A&A...593A.120V, 2018A&A...618A..24V}. It fits individual mode periods and period spacing patterns using asymptotic period spacing predictions based on the traditional approximation of rotation \citep{2005MNRAS.360..465T}. Its output provides model-independent estimates for the near-core rotation, $f_\mathrm{rot}$, and buoyancy period, $\Pi_0$.

    TIC 5536869 (HD 20420) shows a very short pattern containing four consecutive modes and one additional individual mode. We find a moderate rotation $f_\mathrm{rot}=1.88\pm0.07$\,d$^{-1}$ and estimate $\Pi_0=5245\pm1145$\,s. The uncertainty associated with $\Pi_0$ is largely due to the short pattern. Unlike the other two stars the dominant mode does not fit the pattern. TIC 5536869 is a photometric binary and it exhibits multiple strong modes of which some might belong to the second component in the system. Better resolved frequency spectra are needed to disentangle two stars in a pulsating binary from the same pixel in space photometry \citep[e.g.][]{2014MNRAS.441.2515M}.

    TIC 67828699 (HD 23388) has a short but very regular period spacing pattern consisting of four spacings including three consecutive. We estimate $\Pi_0=5150\pm260$\,s and $f_\mathrm{rot}=2.99\pm0.02$\,d$^{-1}$, making it the fastest rotator in our sample with a period spacing. This rotation rate is comparable to stars in NGC\,2516 \citep[][see also below]{Li2024}.

    TIC 385552372 (HD 23375) is a pure \gdor{} star with a short pattern of only two consecutive spacings and an individual mode. It has a fast near-core rotation $f_\mathrm{rot}=2.54\pm0.07$\,d$^{-1}$. We find a slightly lower $\Pi_0=4850\pm1800$\,s compared to the other two stars but with large a uncertainty.

    In addition to these three stars, we found tentative patterns for TIC\,385554826 and TIC\,84331341. Yet, they are too noisy to yield useful parameter estimations. In the case of TIC\,385554826, the consideration of the sector 18 data did not yield additional pattern modes. Despite clearly visible $l=2$, $m=2$ frequencies in several stars, we were not able to identify period spacing patterns of quadrupole modes. Similarly, TIC\,427580304 (one of the two previously known pure \gdor{} stars) does not show a period spacing pattern but exhibits several widely spaced modes. One additional TESS campaign in the ecliptic plane might already be sufficient to enable the detection of more modes in this star, facilitating the construction of a period spacing pattern.

    Given the very short detected patterns, the question arises whether the stars' period spacing patterns are truly smooth or whether their slope could be affected by buoyancy glitches \citep{Cunha2024}. Several properties of the observed patterns indicate that they are truly smooth and the derived asteroseismic properties accurate. Firstly, two out of the three patterns have fitting mode frequencies across several higher radial orders higher than those in the core of the period-spacing pattern. Buoyancy glitches typically manifest in quasi-periodicity in the period spacing pattern along a large range of radial orders \citep{2015ApJS..218...27V}. Secondly, for young stars such as in the Pleiades, a rather smooth period spacing pattern is expected because such stars do not yet exhibit strong chemical gradients near their core causing buoyancy glitches and periodicity in the pattern \citep{SchmidAerts2016}. Finally, as seen in the following Section, the derived asteroseismic near-core rotation rates are in agreement with the estimate based on the dominant mode.

    \subsection{Near-core rotation from dominant gravity modes}

    Rotation does not only affect the slope of the period spacing pattern but also shifts the observed mode frequencies \citep[see Fig. 10 in][]{2017A&A...598A..74P}. The dominant g-mode frequency is linearly dependent on the near-core rotation rate. Exploiting this fact, \cite{Aerts2025} obtained a linear relation based on \emph{Kepler} observations for prograde dipole modes ($l=1$, $m=1$) between the dominant frequency and the near-core rotation rate. As shown by \cite{Li2020} and \cite{Hey2024}, the overwhelming majority of g-mode pulsators exhibit dipole pulsations, making this relation a very useful tool. We deduced $f_\mathrm{rot}$ for all 22 g-mode pulsators without identified period spacing patterns based on the regression formula in Eq. 4 of \cite{Aerts2025} allowing us to estimate the internal rotation rate for the entire sample of g-mode pulsators in the Pleiades. As seen from the values in Table~\ref{tab:astero}, the estimated rotation rates are in good agreement with the measured values from the period-spacing patterns for the three stars with such a pattern. Using this estimate of $f_\mathrm{rot}$ for the remaining 22 pulsators, we complete the picture of internal stellar rotation in this cornerstone open cluster and analyse it in the following Section.

    \section{Rotation on the upper main sequence of the Pleiades}
    \label{sec:rotation}

    Despite being a well surveyed open cluster, the distribution of the rotation rates on the upper main sequence of the Pleiades was until now only studied from $v\sin i$ observations \citep{2020ApJ...901...91T,Bedding2023}. With the sample of near-core rotation rates from g-mode asteroseismology, we are now in a position to explore the stellar rotation in the Pleiades without the inclination effect. To fully study the mass dependence of stellar rotation, we also consider surface rotation measurements from surface modulation and spectroscopic measurements of $v\sin i$.

    For our final combined data set, we collected rotation measurements for 62 out of 105 stars (Sect.~\ref{sec:targetanddata}). These data include asteroseismic near-core rotation rates for \ngdor{} stars, surface rotation rates from spot modulation for 36 stars \citep[and this work]{Rebull2016}, and spectroscopic surface $v\sin i$ measurements for 31 stars \citep{2020ApJ...901...91T}. We recall that $v \sin i$ is a projected value of the overall spectral line broadening, often affected by pulsational velocity broadening leading to both larger and smaller apparent rotational velocities \citep{2014A&A...569A.118A} due to line profile variations. Some stars have multiple measurements. Most intermediate-mass stars rotate rigidly with minimal differential rotation between the core and surface \citep{2018A&A...618A..24V, Li2020, 2021RvMP...93a5001A}; under this assumption we use all rotation rates equally and independent of their probing region.

    \subsection{Photometric surface rotation rates}
    Among cool low-mass stars, light curve modulation from stellar spots (typically due to magnetic activity) is a common phenomenon, which enables the measurement of the true surface rotation period (see \citealt{Rebull2016} for these measurements in the Pleiades). Surface modulation originating from chemical inhomogeneities caused by magnetic fields \citep{2023MNRAS.520..216H} can also be observed in some intermediate-mass field stars. We identified stars with surface modulation and measured their rotation period from the highest peak in the periodogram.

\begin{figure*}
        \includegraphics[width=\textwidth]{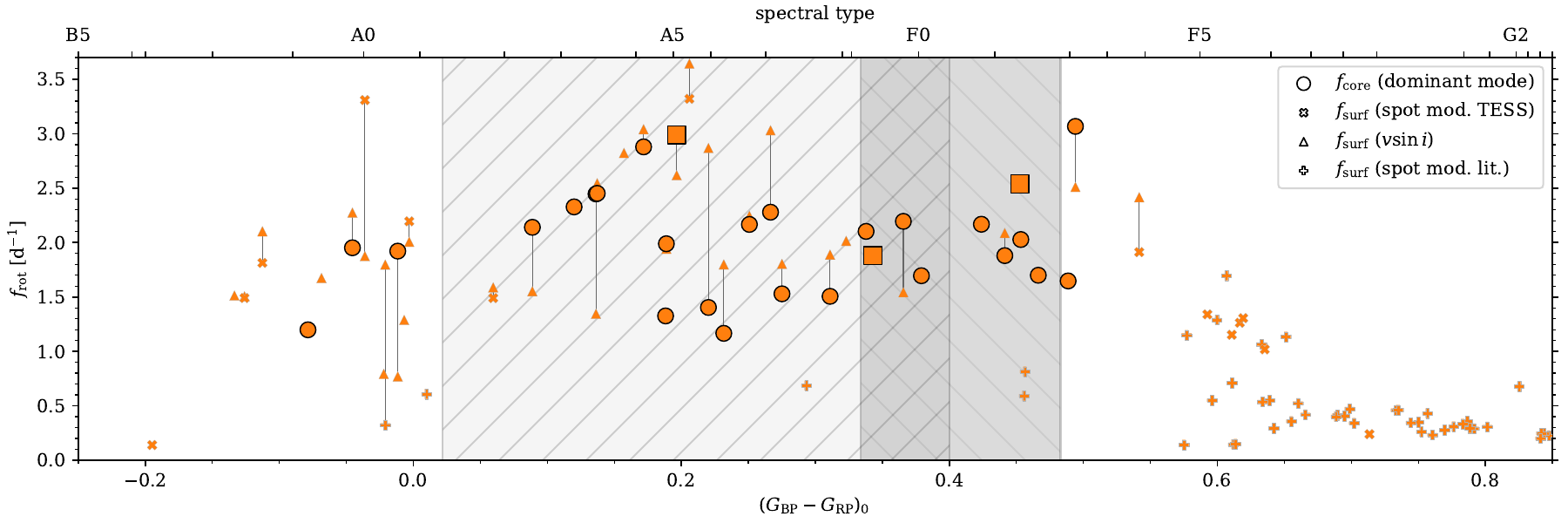}
        \caption{Stellar rotation rates for upper main-sequence stars in the Pleiades. Asteroseismic near-core rotation rates are shown with large symbols, while surface rotation rates are shown with small symbols. The different shapes correspond to the different methods. Squares indicate near-core rotation inferred from period spacing patterns using the traditional approximation of rotation, circles are based on the dominant mode frequency of the g-mode pulsators. Photometric surface rotation rates are marked with plusses for stars from \cite{Rebull2016} and with crosses for new detection in this work. Finally, we show rotation rates estimated from $v\sin i$ measurements in the Pleiades with triangles. Stars with both surface and near-core rotation rates are connected by lines. The grey hatched areas indicate the approximate range of the theoretical \gdor{} \citep[downwards left-to-right,][]{2004A&A...414L..17D} and observational \dscu{} \citep[upwards left-to-right,][]{2019MNRAS.485.2380M} instability strips. The asteroseismic uncertainties are mostly within the symbol sizes and have been omitted for clarity.}
        \label{fig:rotPleiades}
    \end{figure*}

    We find 14 stars among our 105 targets with surface modulation not included in the K2 observations previously analysed by \cite{Rebull2016} (see Table~\ref{tab:photrot}). Six of our new detections are near the Kraft break \citep{1967ApJ...150..551K}, where surface modulation due to magnetic spots is still expected. Among the hotter stars, along the upper main sequence, we find eight stars with probable rotational signals in their frequency spectra. These hotter stars are mostly consistent with the asteroseismic rotation rates (cf. Fig.~\ref{fig:rotPleiades}), while the intermediate-mass rotators from \cite{Rebull2016} are mostly slower rotators below the bulk of the distribution.

    \subsection{Spectroscopic surface rotation rates}

    In order to compare the spectroscopic velocities ($2\pi R f_\mathrm{rot}\sin i$) to the true rotation rates, we estimated the stellar radius, $R$, from the \texttt{MIST} isochrones. For likely single stars, the radius was based on the mean of the radii as obtained from the intrinsic $(G_\mathrm{BP}-G_\mathrm{RP})_0$ colour and from the absolute magnitude $M_G$. For potential photometric binary stars (i.e. stars elevated above the cluster sequence), only the colour was used to avoid systematically overestimating the radius.

    For a rigidly rotating star, the spectroscopic rotation rate should always be smaller than the directly observed rotation rate due to the inclination effect. However, we find five stars to show significantly higher spectroscopic rotation rates compared to our asteroseismic values (with a difference larger than 0.3\,d$^{-1}$). Notably, four of these stars are \dscu{} stars. We conclude that their short period pulsations contribute to spectral line broadening over the exposure time of the spectra, leading to a too large $v\sin i$ measurement. A similar pulsation induced effect could lead to the overestimation in the remaining star \citep{2014A&A...569A.118A}.

    \subsection{(Near-core) Rotation as a function of colour}

    The distribution of (near-core) rotation rates on the upper main sequence in the Pleiades (as shown in Fig.~\ref{fig:rotPleiades}) is centred around 2\,d$^{-1}$ and can be divided into three parts. Firstly, the highest mass stars of spectral type B have a slightly lower internal rotation rate $\sim{}\!1$\,d$^{-1}$ in agreement with \cite{2021RvMP...93a5001A}. Secondly, the bulk of intermediate-mass stars of later spectral types rotate at the aforementioned $\sim{}\!2$\,d$^{-1}$ with some scatter. Thirdly, the transition to cool stars at the Kraft break is indicated by the large number of photometric surface rotation rates declining rapidly with colour.

    We revisit the first group in the following Section and concentrate here on the bulk of upper main sequence stars of spectral types A and F within the instability regions. Despite significant scatter, we find these stars to rotate typically with a rotation rate of $1.0-2.5$\,d$^{-1}$ over the entire considered mass range from early A to mid-F stars. Only three stars fall below this range and all of them exhibit surface modulation in their light curves. The lower envelope is likely not an observational bias for g-mode pulsators; stars without gravity modes (which enable the near-core rotation rate estimation) are either not periodically variable at all or show clear signals of surface modulation with the typical harmonics in the frequency spectrum (Sect.~\ref{sec:nonpuls}). Hence, we would have been able to detect their pulsations. This suggests that some mechanism is at play that limits the rotation rate of intermediate-mass stars between $\sim{}\!1$ and 3\,d$^{-1}$ ($\sim{}50\,\%$ of the critical rotation rate) at young ages. This is in agreement with \cite{2024A&A...685A..21M}, who found that \gdor{} stars have an initial rotation rate distribution up to $\sim\!65\,\%$ of the critical rotation rate.

    For lower mass stars the distribution of rotation rates moves to lower values as the convective envelope thickens, allowing for an efficient dynamo and the magnetic spin down of the stars. The continuity of the rotation rate distribution across the Kraft break highlights that inefficient magnetic breaking is still active in this regime. The gradual transition across the Kraft break reflects the evolution of increasingly thicker convective envelopes. In contrast, \cite{2024ApJ...973...28B} recently found an abrupt transition with a width of only $\delta M=0.1\,M_\sun$. In their study, \cite{2024ApJ...973...28B} selected all F-type stars within 33\,pc but excluded young stars, while the stars in the Pleiades can still be considered young by their definition of young ($\lesssim 150$\,Myr). However, stars more massive than the Sun (i.e. earlier than spectral type G2) that originally resided on the fast rotator sequence \citep[cf.][see also Fig.~\ref{fig:full_frot}]{Fritzewski2020} of the Pleiades have already spun down. The fast rotators observed here are not part of this sequence, but rather stars that will not spin-down in a similar fashion as lower-mass stars as they lack access to efficient magnetic braking. Hence, removing these stars from the sample \citep[as in][]{2024ApJ...973...28B} might introduce a bias as they are fast rotating stars not due to their youth but due to their higher mass. In conclusion, the Kraft break as observed in this coeval population is not a sharp transition but gradual as it is populated by faster rotating stars that do not undergo the same magnetic breaking as lower-mass stars.

    A gap appears between the stars with observed pulsations and stars with surface modulation. In this regime stellar spots are very small due to the lower stellar activity of late-F stars and cannot be measured from TESS photometry due to the noise characteristics of the light curves. Simultaneously, the development of convection in the stellar envelopes causes g-mode oscillations to be evanescent, preventing their detection.

    \section{Comparison to NGC\,2516}
    \label{sec:N2516}
    In order to better understand the rotation rate distribution on the upper main sequence, we wish to compare it to other open clusters. An ideal candidate for such a comparison is the southern open cluster NGC\,2516 which is very similar to the Pleiades. Initially, both clusters were even assumed to be twin clusters \citep{1972ApJ...173...63E}. In combination, the population of low-mass stars of these two open clusters lays out the initial near-ZAMS rotation period distribution for cool stars \citep{Fritzewski2020}. Beyond their identical gyrochronological ages, the lithium-depletion age of both open clusters is also the same \citep{1998MNRAS.300..550J, Bouma2021}. Despite these clear astrophysical evidences of a similar age, NGC\,2516's age has recently been estimated from isochrones as being between 100\,Myr \citep{Li2024} and 750\,Myr \citep{2024A&A...689A..18A, 2025ApJ...979...92W}, while its asteroseismic age was measured to be $132\pm8$\,Myr \citep{Li2025}. For our work, we assume the Pleiades and NGC\,2516 to be (nearly) coeval. We aim to compare the upper main sequence members of the two open clusters using not only their surface properties (as in the case of cool star rotation periods and lithium surface abundances) but their internal asteroseismic properties, namely the buoyancy period and the near-core rotation.

    \subsection{Buoyancy period}

    The buoyancy period, $\Pi_0$, decreases during a star's main sequence life, making it an independent age indicator \citep{2019MNRAS.485.3248M, Fritzewski2024}. Stars of similar mass in the Pleiades and NGC\,2516 should therefore have similar values of $\Pi_0$ under the assumptions that both clusters are of similar age and that its stars have similar mixing properties.

\begin{figure}
        \includegraphics[width=\columnwidth]{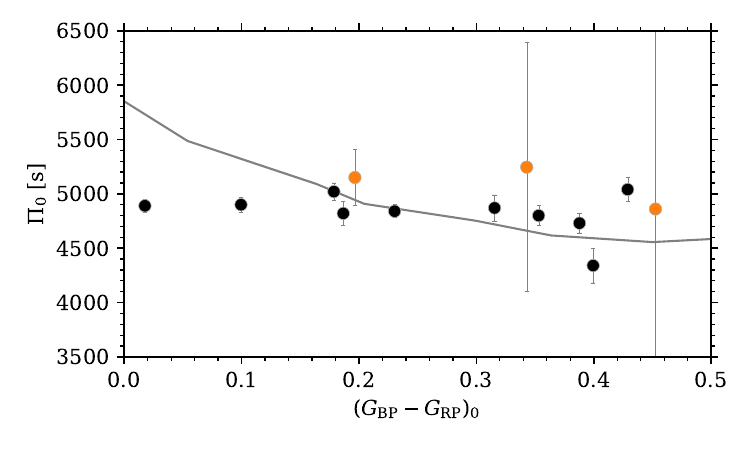}
        \caption{Comparison of the buoyancy period $\Pi_0$ between the Pleiades (orange) and NGC\,2516 (black) shown against the intrinsic \emph{Gaia} colour $(G_\mathrm{BP}-G_\mathrm{RP})_0$. The uncertainties in $\Pi_0$ are much larger for the Pleiades due to the shorter period spacing patterns. The solid line shows a representative isochrone at 130\,Myr from \cite{Li2025}.
        }
        \label{fig:pi0comp}
    \end{figure}

    Figure~\ref{fig:pi0comp} shows a comparison of the derived $\Pi_0$ between the three Pleiades members with period-spacing patterns and the 14 stars in NGC\,2516 from \cite{Li2024}. Most of the members in NGC\,2516 have a buoyancy period close to 5000\,s, a value typical for young main sequence g-mode pulsators. All three Pleiads are consistent with this distribution, considering the large uncertainties on our measurements. These larger uncertainties compared to the stars in NGC\,2516 can be attributed to the short patterns, limiting the precision of the asteroseismic properties derived from the period spacing patterns. NGC\,2516 falls close to the Southern continuous viewing zone of TESS, resulting in much longer photometric time series and more detected modes in the pulsators.

    Beyond a purely empirical comparison, we also show a $\Pi_0$ isochrone at 130\,Myr in Fig.~\ref{fig:pi0comp}. It is derived from the models presented in \cite{Li2025}, in particular from the model with a convective core overshoot $f_\mathrm{ov}=0.025$. From the red-edge of the \gdor{} instability region up to $(G_\mathrm{BP}-G_\mathrm{RP})_0\approx0.2$ the measured buoyancy periods are close to the model. However, the growing trend of $\Pi_0$ with mass is not observed. Similarly, the bluer stars, for which we only have data for NGC\,2516, lie well below the predicted $\Pi_0$ as the observations show no mass-dependence. \cite{Li2025} speculate that this discrepancy between observations and isochrones might be related to the mass-discrepancy observed in binary stars \citep[cf.][]{2020A&A...637A..60T}. Further, the majority of the stars in Fig.~\ref{fig:pi0comp} are fast rotators, while the stars in the \emph{Kepler} field, to which the asteroseismic models are calibrated \citep{2024A&A...685A..21M}, cover a wide range of rotation rates, from very slow to about 60\% of the critical rate \citep{Aerts2026}. Based on two data points, we cannot draw conclusions, but it is of great interest to obtain as many $\Pi_0$ measurements in open clusters in the future to enable an empirical calibration of asteroseismic models over a large mass, age, and rotation range.

    The large uncertainties prevent a strong statement on the coevality of the two clusters, but we can at least say that the current asteroseismic data are consistent with a coeval state. With only short patterns detected, a detailed asteroseismic modelling following the seismic age-dating of NGC\,2516 \citep{Li2025} is impractical, preventing a precise asteroseismic quantification of the age difference between the two open clusters for now.

    \subsection{Do stars in the Pleiades rotate slower compared to NGC\,2516?}

    The internal angular momentum transport and evolution of early-type stars is still not understood \citep[see][for a review]{2019ARA&A..57...35A}. To progress our understanding of these fundamental processes, large samples of asteroseismic rotation rates are required. From one such sample, \cite{Aerts2025} recently found near-core rotation rates to decrease during main sequence evolution, implying efficient angular momentum transport within the radiative envelope. Samples of near-core rotation rates from open clusters can act as calibrators of rotating asteroseismic models. For cluster members the age can be measured independently, constraining the stellar models. Here, we compare the two nearly coeval open clusters Pleiades and NGC\,2516 to provide insights into cluster-to-cluster differences while establishing a (near-) initial rotation distribution for intermediate-mass stars.

    \begin{figure*}
        \includegraphics[width=\textwidth]{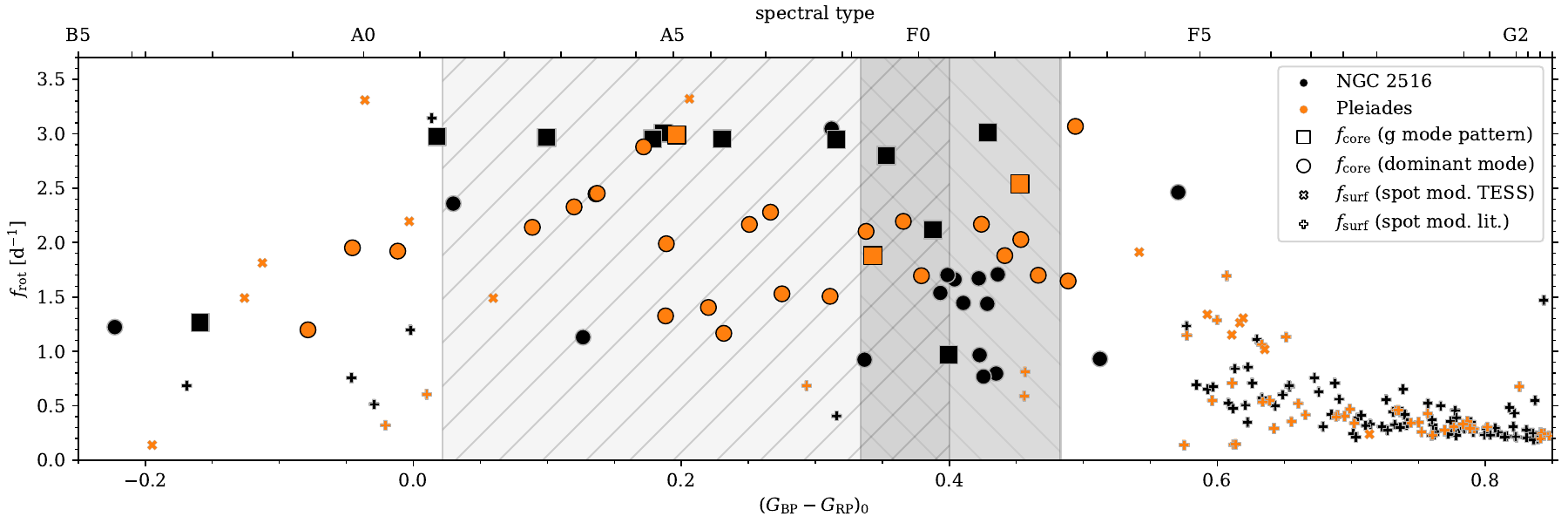}
        \caption{Stellar rotation rate distribution for stars in the Pleiades (orange) and NGC\,2516 (black) similar to Fig.~\ref{fig:rotPleiades}. Here, $v\sin i$ based rotation rates have been omitted for clarity.
        }
        \label{fig:rotationcompare}
    \end{figure*}

    This rotation distribution is laid-out in Fig.~\ref{fig:rotationcompare}, which compares the observed rotation rates in the Pleiades and NGC\,2516. The first notable difference between the two clusters is the much larger degree of scatter present in the Pleiades. In NGC\,2516, nearly all A-type stars rotate near 3\,d$^{-1}$, while the Pleiades members are mostly distributed between 1\,d$^{-1}$ and 2.5\,d$^{-1}$. Still, the fastest rotating Pleiad has also a rotation rate of $~\!3$\,d$^{-1}$.

    The large spread in rotation rates for the Pleiades compared to NGC\,2516 over this large mass range is of interest. Under the assumption of a constant angular momentum evolution (i.e. no angular momentum loss) for intermediate-mass stars \citep{2013MNRAS.429.1027A}, stars of different rotation rates but similar age and mass must have had different initial angular momentum contents. Given the large spread in rotation rates of field stars \citep{2024A&A...685A..21M, Aerts2025}, NGC\,2516 seems to be the outlier here.
    In the classical \gdor{} instability strip the roles of the two clusters are reversed. Here, we find a larger spread in NGC\,2516 and the transition to the cooler stars through the Kraft break appears less smooth compared to the Pleiades, although we are dealing with few stars.
    At the moment, we cannot offer a firm explanation for the different angular momentum content in these two, otherwise very comparable, open clusters. NGC\,2516 is located in the southern PLATO field of view \citep{2025A&A...694A.313N} and future observations might provide more insights into whether this difference is physical or an observational bias.

    \subsection{Discussion}

    \cite{Li2024} found that nearly all intermediate-mass stars in NGC\,2516 rotate with $f_\mathrm{rot}=3$\,d$^{-1}$ corresponding to about 50\,\% of their critical velocity. The rotation rates found in the Pleiades top-out at a similar value, both for the asteroseismic near-core rotation as well as the spectroscopic $v\sin i$ (cf. Fig.~\ref{fig:rotPleiades}). Although we find many slower rotating stars in the Pleiades, this maximum rotation frequency over such a large mass range is remarkable.
    It is in agreement with the large sample of g-mode pulsators in the Galaxy, whose rotation rates range from zero up to about 70\% of the critical rate \citep{Aerts2026}.
    A-type field stars rotating at 80\,\% critical velocity are still physical and have been observed \citep[e.g. Altair,][]{2020A&A...633A..78B} but it concerns very few stars pulsating in pressure modes. At the moment, we can only speculate about the origin of this upper limit for the g-mode pulsators. It can either be an observational bias or astrophysical.

    Firstly, we explore potential biases introduced by the methodology or the observations. Although we tried to identify all g-mode pulsators in Pleiades, we might have missed some of the fastest rotators as the dominant gravity mode moves to shorter periods (higher frequencies) for faster rotating stars. These short periods blend with \dscu{} pulsations in the frequency spectrum. In addition, the photometry of some Pleiades members is highly contaminated by the naked-eye stars in the cluster core reducing the number of members with suitable photometry.

    Another possibility for this seemingly strict upper limit in near-core rotation frequency might be found in the mode excitation.  Very few g-mode pulsators with $f_\mathrm{rot}\gg3$\,d$^{-1}$ (or critical rotational velocity above 70\,\%) have been found in the mass regime of the AF-type pulsators, which are destabilised by the flux blocking mechanism \citep{Dupret2005}. The fastest rotators in the \emph{Kepler} field \citep{Li2020} are comparable to the stars in the Pleiades and NGC\,2516, which seems to indicate that the excitation and propagation of gravito-inertial modes under fast rotation becomes ineffective in stars with a thin convective envelope rotating close to the critical rate. Several stars with a mass above 2.5\,M$_\odot$ do reveal faster rotation and gravity modes \citep[][Fig.\,5]{Aerts2026}. These higher-mass pulsators are excited by the $\kappa$ mechanism active in a purely radiative envelope, which seems to allow for gravito-inertial modes in stars rotating up to the critical rate \citep{2021NatAs...5..715P}.

    In the rotating pre-main sequence stellar models presented by \cite{2017A&A...602A..17H} a similar constant rotation rate is reached over a large mass range (covering A and F stars) at ZAMS. However, this value is ultimately set by the author's chosen angular momentum accretion history and in their earlier work \citep{2013A&A...557A.112H}, intermediate-mass stars reached critical velocity. Observationally, stars of very different rotation rates, including near-critical rotation are known. The mechanism responsible for setting the angular momentum accretion history are still unknown. Studies of very young g-mode pulsators (e.g. Wadhwa \& Fritzewski, in prep.) have the potential to probe some of the aspects of the earliest angular momentum evolution.

    Similarly to the upper limit, a lower envelope of rotation rates is visible near 1\,d$^{-1}$ in both clusters. Given that this lower envelope is present in both the data of NGC\,2516 and the Pleiades, it should not be an observational bias. The former cluster is located near the Southern continuous viewing zone of TESS, so that frequencies close to 1\,d$^{-1}$ could be resolved. The only rotation frequencies measured for slower stars arise from surface modulation\footnote{Periodic light curve modulation from surface spots leads to harmonics in the frequency spectrum, while gravity modes typically manifest as a (un-)resolved comb. These difference enable the clear separation between pulsations and rotational signals.}.

    Slowly rotating stars, both from asteroseismic near-core and photometric surface rotation rates, are found among the B stars in both clusters. As seen from the CMD in Fig.~\ref{fig:CMD} (and Fig. 8 in \citealt{Li2024}), these stars are evolved the furthest and their radii are significantly larger than those of ZAMS stars with the same mass. Their slower near-core rotation implies that angular momentum transport is still very effective on the second half of the main sequence in line with the findings in \citet{Aerts2025} and \citet{Aerts2026}. 

    Detailed modelling of both clusters is needed to draw stronger conclusions on the differences and similarities of the intermediate-mass stars between these two clusters. A first step has already been taken by \cite{Li2025}, who derived an asteroseismic age for NGC\,2516 that agrees with the lithium depletion age of the Pleiades reaffirming their similar ages. The small number of short period spacing patterns in this work preclude similar analysis; a different approach is needed to fully explore the asteroseismic potential of the Pleiades with the currently available data.

    \section{Conclusions}
    \label{sec:conclusions}

    We present a comprehensive survey of gravity-mode pulsators in the Pleiades open cluster based on TESS data. The entire upper main sequence ($M_\star\gtrsim1.3\,M_\sun$) of the open cluster is populated with g-mode pulsators, making it a very rich population of pulsators. Taking into account the previously studied pressure mode pulsators (\dscu{} stars, \citealt{Bedding2023}), we find a large fraction of hybrid pulsators. Overall, we identify \ngdor{} g-mode pulsators of which \nhybrid{} are hybrid pulsators making the Pleiades an ideal test bed for future studies of hybrid pulsators.

    Three of the g-mode pulsators show short period spacing patterns that could be modelled with the traditional approximation of rotation to obtain near-core rotation rates and asymptotic buoyancy periods. Many unambiguous g-mode pulsators do not show patterns due to the relatively short time series, preventing the detection of low amplitude modes.

    For the other 25 g-mode pulsators, we derived their near-core rotation rate from their dominant prograde dipole mode \citep{Aerts2025} and find a wide range of rotation rates over the A- and early F-type spectral range. These stars have rotation rates randomly distributed between 1 and 3\,d$^{-1}$ with no clear colour- (or mass-) dependence.

    This is a more diverse result compared to what was observed in NGC\,2516 by \cite{Li2024}, who found that the vast majority of A-type members rotate at 3\,d$^{-1}$ while only the early F-type stars populate the region between 1 and 3\,d$^{-1}$. The reasons for these different angular momentum contents of stars in these two open clusters are elusive at the moment. Future observations of NGC\,2516 from PLATO might reveal more details about the g-mode pulsators in NGC\,2516.

    The derived buoyancy periods are consistent with a similar age of the Pleiades and NGC\,2516 but large uncertainties inhibit us from placing a strong constraint. Detailed asteroseismic modelling would be required to quantify the agreement of the ages and compare it to the cool stars in both clusters. A first step has already been taken for NGC\,2516 by \cite{Li2025}. However, asteroseismic modelling of the Pleiades pulsators is more challenging because of the few identified pulsation modes.

    Our search for g-mode pulsators in the Pleiades uncovered a large and novel population in this very well studied open cluster. This initial study lays the groundwork for future asteroseismic analysis of this cornerstone open cluster. It is reaffirming the position of the Pleiades as the prime benchmark object in stellar astrophysics.

    \begin{acknowledgements}
        We are grateful to the anonymous referee for their insightful and helpful report. We thank Simon Murphy for comments on the initial submission.
        The research leading to these results has received financial support from the Flemish Government under the long-term structural Methusalem funding program by means of the project SOUL: Stellar evolution in full glory, grant METH/24/012 at KU Leuven.
        CA acknowledges financial support from the European Research Council (ERC) under the Horizon Europe programme (Synergy Grant agreement N$^\circ$101071505: 4D-STAR). While partially funded by the European Union, views and opinions expressed are however those of the author(s) only and do not necessarily reflect those of the European Union or the European Research Council. Neither the European Union nor the granting authority can be held responsible for them.
        This research has made use of NASA's Astrophysics Data System Bibliographic Services and of the SIMBAD database and the VizieR catalogue access tool, operated at CDS, Strasbourg, France.
        This work has made use of data from the European Space Agency (ESA) mission \emph{Gaia} (\url{https://www.cosmos.esa.int/gaia}), processed by the \emph{Gaia} Data Processing and Analysis Consortium (DPAC, \url{https://www.cosmos.esa.int/web/gaia/dpac/consortium}). Funding for the DPAC has been provided by national institutions, in particular the institutions participating in the \emph{Gaia} Multilateral Agreement.
        This paper includes data collected by the TESS mission, which are publicly available from the Mikulski Archive for Space Telescopes (MAST).
        \newline
        \textbf{Software:}
        This research made use of \texttt{Astropy}, a community-developed core \texttt{Python} package for Astronomy \citep{2013A&A...558A..33A} and \texttt{Lightkurve}, a \texttt{Python} package for Kepler and TESS data analysis \citep{2018ascl.soft12013L}.
        This work made use of \texttt{Topcat} \citep{2005ASPC..347...29T}.
        This research made use of the following \texttt{Python} packages:
        \texttt{astroquery} \citep{2019AJ....157...98G};
        \texttt{IPython} \citep{ipython};
        \texttt{MatPlotLib} \citep{Hunter:2007};
        \texttt{NumPy} \citep{numpy};
        \texttt{Pandas} \citep{pandas};
        \texttt{SciPy} \citep{scipy}

    \end{acknowledgements}

    \bibliographystyle{aa} 
    \bibliography{pleiades.bib} 

    \begin{appendix}

    \section{Data tables}
    \begin{table*}
    \caption{Asteroseismic near-core rotation frequencies and buoyancy periods of g-mode pulsators in the Pleiades.}
    \label{tab:astero}
    \resizebox{\textwidth}{!}{
    \begin{tabular}{rlrrrrrrrrrrr}
        \hline
        \hline
        TIC & Name & \emph{Gaia} DR3 source\_id & RAdeg & DEdeg & Gmag & $(G_\mathrm{BP}-G_\mathrm{RP})_0$ & $\Pi_0$ & $\delta\Pi_0$ & $f_\mathrm{rot}$ & $\delta f_\mathrm{rot}$ & $f_\mathrm{rot}$ ($f_\mathrm{dom}$) & $\delta f_\mathrm{rot}$ ($f_\mathrm{dom}$)\\
         & & & ($\deg$) & ($\deg$) & (mag) & (mag) & (s) & (s) & (d$^{-1}$) & (d$^{-1}$) & (d$^{-1}$) & (d$^{-1}$)\\
         \hline
        5536869 & HD 20420 & 62413988007668352 & 49.45744 & 22.83199 & 7.579 & 0.343 & 5245 & 1145 & 1.880 & 0.073 & 1.93 & 0.04\\
        26126738 & HD  22146 & 67799189799869312 & 53.74454 & 23.52995 & 8.786 & 0.453 & \dots & \dots & \dots & \dots & 2.03 & 0.04\\
        35156298 & HD 23763 & 66725379258016768 & 57.12550 & 24.34532 & 6.940 & 0.136 & \dots & \dots & \dots & \dots & 2.45 & 0.05\\
        46476992 & HD 21744 & 69335619861034752 & 52.81658 & 25.25526 & 8.088 & 0.275 & \dots & \dots & \dots & \dots & 1.53 & 0.02\\
        61139371 & HD 23512 & 65008698009126528 & 56.64259 & 23.62382 & 8.040 & 0.494 & \dots & \dots & \dots & \dots & 3.07 & 0.07\\
        67788829 & HD  23290 & 63378981257953152 & 56.18699 & 20.74780 & 8.633 & 0.466 & \dots & \dots & \dots & \dots & 1.70 & 0.03\\
        67828699 & HD 23388 & 63502259702709888 & 56.38338 & 21.24648 & 7.728 & 0.197 & 5150 & 260 & 2.990 & 0.020 & 3.07 & 0.07\\
        84331341 & HD 24132 & 66657415696297344 & 57.86349 & 24.51845 & 8.770 & 0.489 & \dots & \dots & \dots & \dots & 1.65 & 0.03\\
        113956708 & HD  22578 & 64679840952155264 & 54.66977 & 22.65941 & 6.702 & -0.045 & \dots & \dots & \dots & \dots & 1.95 & 0.04\\
        125736946 & HD 23489 & 65231486552803328 & 56.61376 & 24.25480 & 7.353 & 0.089 & \dots & \dots & \dots & \dots & 2.14 & 0.04\\
        149980785 & HD 24899 & 65776736943479808 & 59.58717 & 24.08092 & 7.198 & -0.012 & \dots & \dots & \dots & \dots & 1.92 & 0.03\\
        346626099 & HD 23886 & 66555985749157760 & 57.35837 & 24.24750 & 7.956 & 0.188 & \dots & \dots & \dots & \dots & 1.33 & 0.02\\
        346626171 & *  28 Tau & 66529975427235712 & 57.29683 & 24.13650 & 5.203 & -0.079 & \dots & \dots & \dots & \dots & 1.20 & 0.01\\
        346626294 & HD 23863 & 66506331628024832 & 57.30088 & 23.88659 & 8.102 & 0.251 & \dots & \dots & \dots & \dots & 2.17 & 0.04\\
        348639016 & HD 23246 & 65292234570088064 & 56.10726 & 24.39450 & 8.122 & 0.338 & \dots & \dots & \dots & \dots & 2.10 & 0.04\\
        385509282 & HD 23325 & 65275501377570944 & 56.27734 & 24.26331 & 8.548 & 0.424 & \dots & \dots & \dots & \dots & 2.17 & 0.04\\
        385552144 & HD 23361 & 65222209423382912 & 56.35903 & 24.03494 & 8.016 & 0.267 & \dots & \dots & \dots & \dots & 2.28 & 0.05\\
        385552372 & HD  23375 & 65296357738731008 & 56.39369 & 24.46311 & 8.550 & 0.453 & 4860 & 1800 & 2.540 & 0.070 & 2.28 & 0.05\\
        385553714 & HD  23323 & 70347033119553408 & 56.31262 & 26.89125 & 8.549 & 0.311 & \dots & \dots & \dots & \dots & 1.51 & 0.02\\
        385554826 & HD 23336 & 71729531553625984 & 56.36485 & 28.66848 & 7.395 & 0.120 & \dots & \dots & \dots & \dots & 2.33 & 0.05\\
        385558439 & HD 23430 & 69872044096705664 & 56.49652 & 25.39839 & 8.017 & 0.232 & \dots & \dots & \dots & \dots & 1.17 & 0.01\\
        385589694 & HD 23409 & 65221487868870784 & 56.46525 & 24.03868 & 7.834 & 0.220 & \dots & \dots & \dots & \dots & 1.40 & 0.02\\
        405461432 & \dots & 117672075163287680 & 52.23637 & 26.30842 & 8.331 & 0.379 & \dots & \dots & \dots & \dots & 1.70 & 0.03\\
        405483425 & HD 23155 & 69877988331531904 & 55.93010 & 25.08049 & 7.513 & 0.137 & \dots & \dots & \dots & \dots & 2.45 & 0.05\\
        405484093 & V1210 Tau & 65212107660378880 & 56.76766 & 23.99501 & 8.332 & 0.366 & \dots & \dots & \dots & \dots & 2.19 & 0.04\\
        427580304 & V* V1225 Tau & 68593346433465856 & 54.96327 & 25.19466 & 8.754 & 0.441 & \dots & \dots & \dots & \dots & 1.88 & 0.03\\
        440681425 & V650 Tau & 65007083101413888 & 56.86189 & 23.67813 & 7.747 & 0.172 & \dots & \dots & \dots & \dots & 2.88 & 0.07\\
        440691730 & HD  23852 & 64114245300877184 & 57.29700 & 22.60926 & 7.706 & 0.189 & \dots & \dots & \dots & \dots & 1.99 & 0.04\\
        \hline
        \end{tabular}
}
    \end{table*}

    \begin{table}
    \caption{New photometric rotation periods for Pleiades members.}
    \label{tab:photrot}
        \begin{tabular}{rrrrr}
        \hline
        \hline
        TIC & RAdeg & DEdeg & $P_\mathrm{rot}$ & $(G_\mathrm{BP}-G_\mathrm{RP})_0$\\
        & ($\deg$) & ($\deg$) & (d) & (mag)\\
        \hline
        14177821 & 59.50716 & 20.67658 & 0.868 & 0.611\\
        15900772 & 62.23087 & 20.38578 & 0.792 & 0.617\\
        46538779 & 53.53057 & 24.34424 & 0.982 & 0.635\\
        46629595 & 54.03535 & 27.34276 & 4.182 & 0.714\\
        67830321 & 56.41634 & 22.69427 & 0.301 & 0.206\\
        113981021 & 54.80510 & 21.84305 & 0.671 & 0.060\\
        348890811 & 61.65183 & 27.59967 & 7.156 & -0.195\\
        353928999 & 55.01295 & 27.74032 & 0.766 & 0.619\\
        385508971 & 56.29068 & 24.83906 & 0.671 & -0.126\\
        405484171 & 56.83776 & 24.11607 & 0.302 & -0.036\\
        427735820 & 58.22292 & 24.71553 & 0.746 & 0.593\\
        440686442 & 57.08683 & 23.42104 & 0.551 & -0.113\\
        440691379 & 57.38645 & 23.38020 & 0.523 & 0.542\\
        440691760 & 57.40917 & 22.53329 & 0.456 & -0.003\\
        \hline
        \end{tabular}
    \end{table}

    \FloatBarrier

    \section{Supporting figures}
     \label{app:figures}

    \begin{figure*}
    \includegraphics[width=\textwidth]{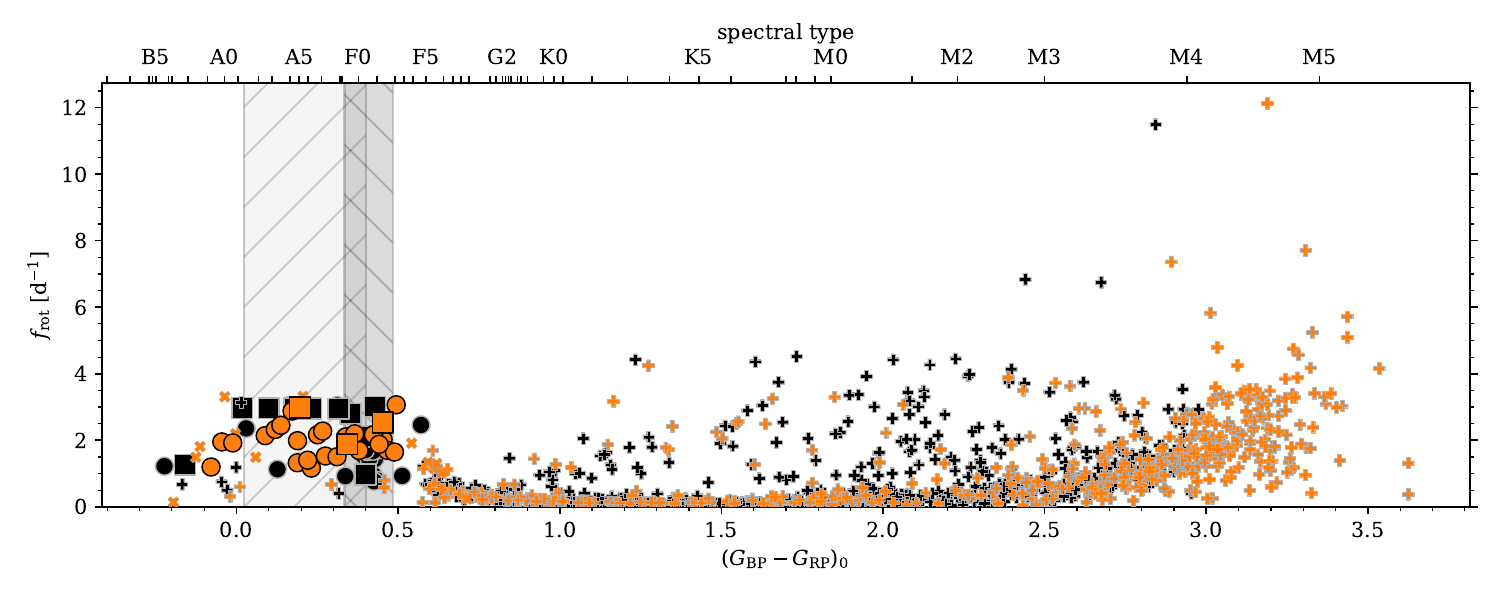}
    \caption{Rotation distribution for all stars with rotation measurements in the Pleiades (orange) and NGC\,2516 similar to Fig.~\ref{fig:rotPleiades}.
    }
    \label{fig:full_frot}
    \end{figure*}

    \begin{figure*}
        \includegraphics[width=\textwidth]{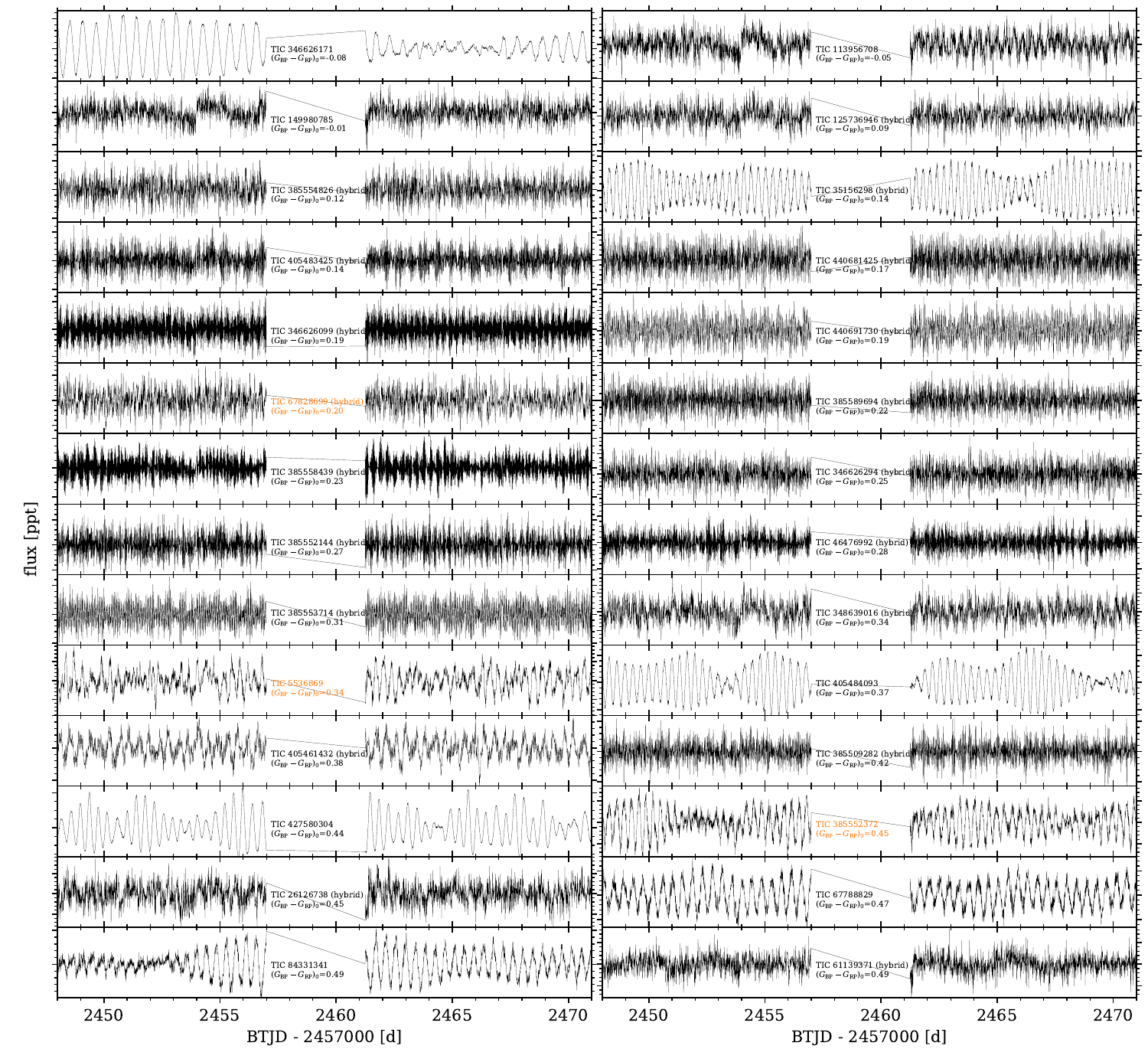}
        \caption{TESS Sector 42 light curves for all identified g-mode pulsators. The asteroseismic analysis is based on sectors $42-44$ and $70-71$ but only one sector is shown here for simplicity.}
        \label{fig:lc_full}
    \end{figure*}

    \begin{figure*}
        \includegraphics[width=\textwidth]{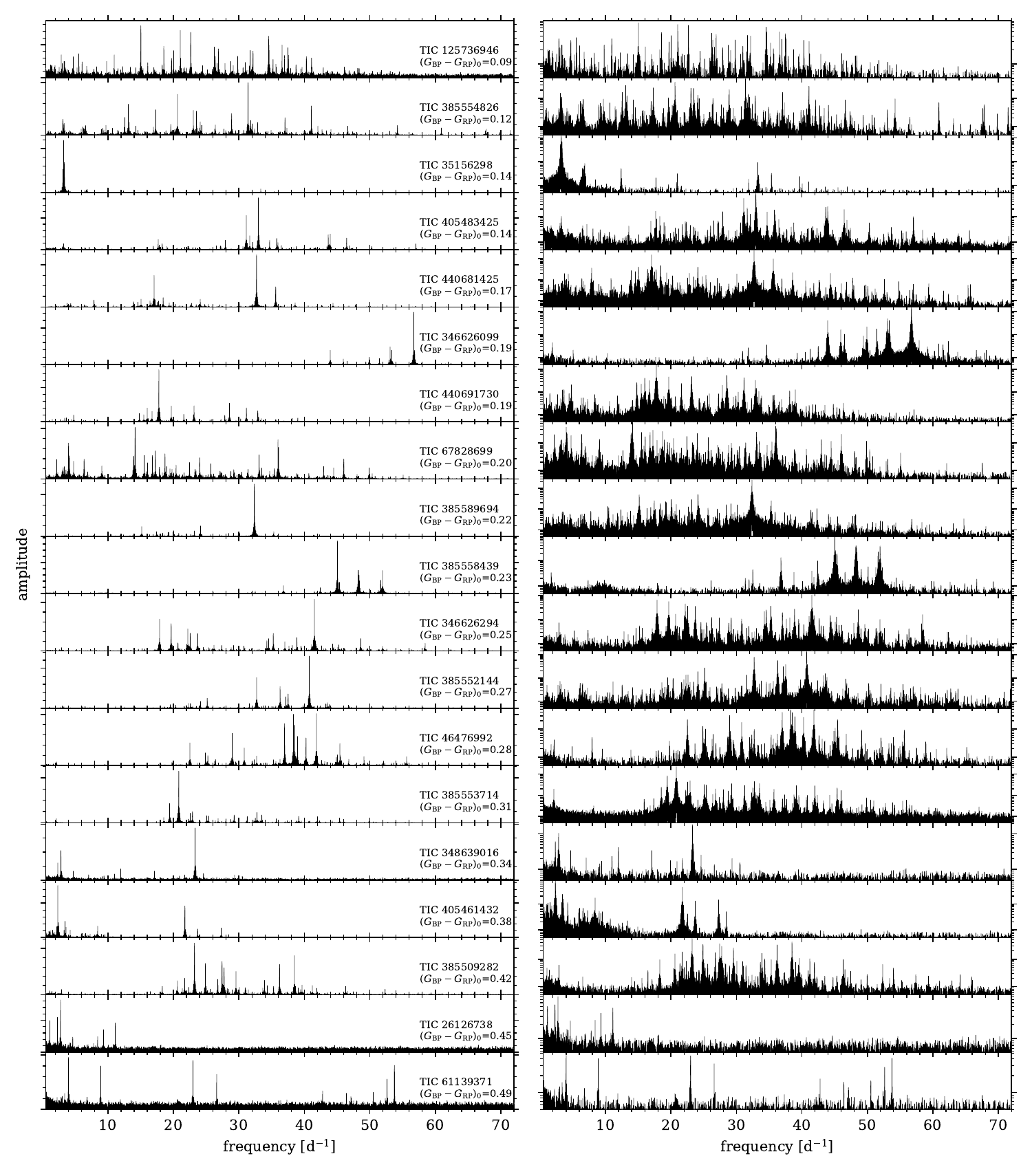}
        \caption{Full frequency spectra for all hybrid pulsators. The left column shows the amplitudes on a linear scale, while the right column is shown in logarithmic scale to highlight the low amplitude modes.}
        \label{fig:pg_full_hybrids}
    \end{figure*}

    \end{appendix}

\end{document}